\documentstyle[times,pramana,epsf,floats]{ias}
\begin{document}

\title{On deglobalization in QCD}

\author{Mrinal Dasgupta}
\address{Theory Division, CERN, 
CH-1211 Geneva 23, Switzerland}
\abstract{
The recent discovery and resummation of a class of single-logarithmic effects (non-global logs), has a significant impact on several QCD observables ranging from the classic Sterman-Weinberg jet definition to currently studied 
event shapes and rapidity gap observables. 
The discovery of the above effects overturns, for example, the common wisdom that hadronic energy flow in limited inter-jet regions is dictated {\it{primarily}} by the colour flow of the underlying hard partonic subprocess.
We discuss some features of non global logs and the rapid progress being made in estimating and controlling such corrections.}
\maketitle

\section{Introduction}
Techniques have been in place for well over a decade to deal with the 
resummation of large logarithms arising in perturbative QCD predictions. 
Large logarithms arise in the first place due to studying problems involving
disparate scales where typically transverse momenta of real quanta are restricted below a value $\epsilon Q$ ($\epsilon \ll 1$) 
where $Q$ is a scale that sets the hardness of the process (e.g the centre-of--mass energy in $e^{+}e^{-}$ annihilation).  Virtual quanta are unrestricted (not affecting the final state measurement) and real-virtual logarithmic divergence cancellations leave behind 
logs $\alpha_s^{n} \ln^{m}{1/\epsilon}$ where $m \leq 2n$. Such logarithms inevitably spoil the convergence of fixed order perturbation theory and must in general be resummed to all orders.

It was generally believed that the technology to perform such resummations 
was already in place, at least to the extent of accounting for single logarithmic ($m=n$) 
and more singular contributions ($m>n$) to many observables.  
However it has recently been discovered \cite{DS1,DS2} that 
a hitherto unforeseen problem plagues single-log contributions to a wide 
variety of observables, previously believed to be well understood at the single-log level. 

Broadly speaking the problem appears for any observable which has a kinematic 
dependence on emissions in only a slice of phase space, 
hence such observables are termed {\it{non global}}.
Since most experimental measurements do place phase space restrictions on the final state one can expect the class of non-global observables to be fairly widespread. 
This is indeed the case and one finds in this category event shapes, rapidity gap observables, jet fractions defined through cone type algorithms, accompanying energy flows in multijet events amongst others. 

For such observables standard methods based on using angular ordering to reduce the problem to that of independent multiple soft emission by jets \footnote{We use the term loosely, we mean really the hard initiating parton.} 
are insufficient at the single-log level.
\section{Non global logs}
To take the discussion further we shall use as illustration the Sterman-Weinberg two jet fraction for $e^{+}e^{-}$ 
annihilation \cite{SW}.
This is defined as the probability of having all but some small fraction $\epsilon$ of hadronic energy flowing inside 
two narrow cones of half-angle delta ($\epsilon,\delta \ll 1$).
One is then left with the task of resumming large logarithms in 
$\epsilon$ and $\delta$ 
in order to recover a meaningful perturbative answer. 

\begin{figure}[htbp]
\epsfxsize=8cm
\centerline{\epsfbox{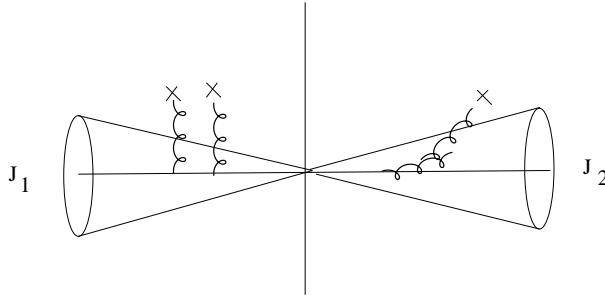}}
\caption{Independent emission (Sudakov) from $J_1$ alongwith correlated (non-global) contribution from $J_2$}
\label{fig:SW}
\end{figure}

To address the above issue one can use the resummation techniques in place prior to the discovery of non-global observables. 
These use angular ordering (independent emission approach) to obtain a 
result  
proportional to a product of Sudakov form factors (one for each jet), with the jet form factor being (omitting for simplicity the running of the coupling):
\begin{equation}
\label{jet}
F_J \sim e^{-2 C_F \alpha_s/\pi \left(\ln \frac{1}{2\epsilon}-3/4\right) \ln \frac{1}{\delta}}.
\end{equation}

We point out that the technique that leads to the above result is essentially based on a 
straightforward exponentiation of the single gluon emission probability. 
The dynamics of subsequent gluon branching is neglected (at single log level) 
due to collinear and infrared safety of the observable in question, apart from the fact that gluon branching contributes to the running of the coupling in the single emission probability (ie is a dressing of the ``bare'' single emission): 
\begin{equation}
d{\mathcal{P}} = C_F \frac{\alpha_s}{\pi} d\eta dk_t \longrightarrow C_F \frac{\alpha_s(k_t)}{\pi} d\eta \frac{dk_t}{k_t}
\end{equation}
In the above, $k_t$ is the transverse momentum and $\eta$ the rapidity of the emission relative to the emitting jet. To accomodate hard collinear splittings and the resulting single-logs, one uses essentially the soft spectrum above but restores the full quark $\to$ quark, gluon splitting function, rather than merely the pole part given by the $1/k_t$ singularity.

Although there are several complications of detail which might depend on the  
case at hand, the above approach is at the heart of most, if not all, 
resummations of large logs in QCD observables. 
This includes angular ordered parton 
showers that are embedded in QCD event generators to simulate whole events, resummations of jet shape variables, Drell Yan vector boson $p_t$ and many others.
It has only recently been demonstrated that there are a  number of observables where such a straightforward picture, although employed in the past, 
is incomplete at single-log level.

Infact for the independent emission result to be valid one should show that there are no single logs arising from a contribution where a primary gluon (one generated directly by the hard jets) branches into two offspring, one of which is much softer than the other. We have already noted that the independent emission ansatz {\it{does}} accommodate collinear non-soft branchings, via the running coupling. It assumes the soft branchings to not contribute due to infrared safety of the observable. 
This would indeed be the case if the observable always triggered both the offspring 
of such a branching -- in other words for a global observable.

However the Sterman-Weinberg two jet fraction (and other non-global observables) is {\it{not}} inclusive over gluon branchings, as we will demonstrate, and consequently  independent emission does not generate the $\alpha_s^n \ln^n\frac{1}{\epsilon}$ terms that would be present in the full $n^{th}$ order 
calculation. For $\epsilon \sim \delta$ these missing pieces would be potentially as important as the single-logs in $\delta$, resummed by the independent emission approach.

To illustrate the above more clearly it helps to consider 
the matrix element squared for ordered two soft gluon emission, 
$W_{12}^{3a}$ where $1$ and $2$ label the hard partons ($q \bar{q}$ pair)
and $3$ and $a$ the soft gluons with $\omega_3 \gg \omega_a$. 
Let us consider specifically the configuration where the softer parton $a$ flies outside the jet cones while $3$ is contained within $J_2$ (see fig.~\ref{fig:SW}). 
The contribution from this configuration goes as 
\begin{equation}
\sigma^2_{{\mathrm{ng}}} = \alpha_s^2\int_{3 \in J_1/J_2} \frac{d\omega_3}{\omega_3} d^2 \vec{n_3} \int_{a \notin J_1/J_2} \frac{d\omega_a}{\omega_a} d^2 \vec{n_a} W_{12}^{3a}\left(\Theta(\epsilon-\omega_a))-1 \right ) 
\end{equation}
The step function represents the constraint on real emission outside the jet cones whilst the $-1$ accounts for virtual corrections. 
There is clearly a mismatch between real and virtual pieces which is not accounted for by independent emission formulae which assume the above contribution to vanish. In the ``inclusive' configurations where both soft partons 
fly into or outside jet cones there is no contribution and the independent emission picture is restored. 

An explicit computation (using the $C_F C_A$ (correlated rather than independent gluon emission) part of $W_{12}^{3a}$ yields, assuming very narrow cones, 
($\delta \ll 1$)  
\begin{equation}
\sigma^2_{{\mathrm{ng}}} \sim -C_F C_A \frac{2\pi^2}{3} \alpha_s^2 \ln^2 \frac{1}{\epsilon}.
\end{equation}
This demonstrates that the Sterman-Weinberg jet fraction is indeed 
a non global observable. In general it will receive single log enhancements $\alpha_s^n \ln^n \frac{1}{\epsilon}$ for $n \geq 2$ from multisoft contributions not included in the independent emission terms. 
At $n^{\mathrm{th}}$ order we will have the non global log arising due to coherent emission of a single softest gluon outside the jet cone 
from a clump of $n-1$ energy ordered soft gluons inside the jet cones. The energy ordering of the $n$ soft gluons produces the logarithm to the $n^{\mathrm{th}}$ power. 
Note that non-global logs are to do with multiple 
gluon emission at relative wide-angles with one another and the emitting hard jets, 
there are no collinear enhancements in this piece. 
\section{Resummation}
The resummation of non global logarithms is rather involved due to the 
following features 
\begin{itemize}
\item Complex colour algebra involved in computing the squared matrix element for the correlated multisoft configuration
\item Complex 3-D geometry of large angle gluon ensemble that does not lend itself to analytical computations
\end{itemize}

\begin{figure}[htbp]
\epsfxsize=8cm
\centerline{\epsfbox{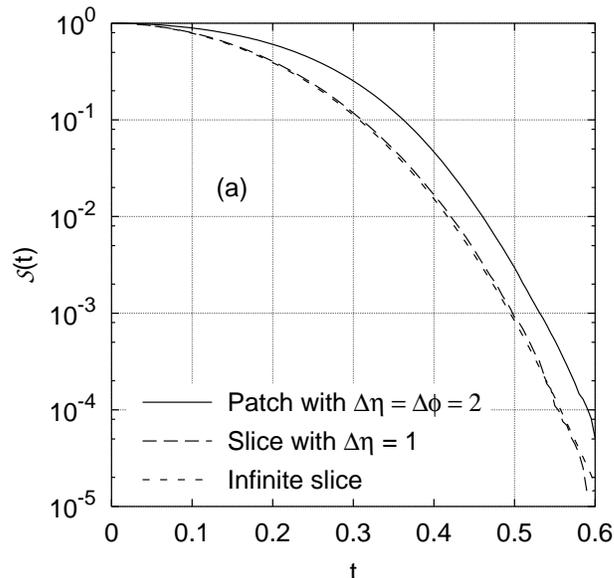}}
\caption{Resummed non-global contribution for energy flow into different regions in the $\eta,\phi$ plane.}
\label{fig:food}
\end{figure}
The first problem can be largely dealt with by resorting to the large $N_c$ limit. In this limit the emission pattern simplifies to a sum over colour dipoles. The evolution of the dipole system is then implemented in a Monte Carlo algorithm such that emission outside the jet cones (or in general in any particular region of interest) is constrained as required. Hence a numerical answer can be readily obtained in the large $N_c$ limit \cite{DS1,DS2}. 

In fact it is possible to write down a non-linear evolution equation for the dipole system \cite{BMS} without implementing it in a Monte Carlo algorithm, but this equation can only be solved numerically so far (at very small values of the energy $\epsilon$) 
and the solution agrees with that provided by the completely equivalent 
Monte Carlo evolution. 
What both the Monte-Carlo and the evolution equation 
leave out are terms of relative order $\left(\frac{1}{N_c^2} \right )$ and hence one can expect a difference at the 10\% level between answers computed currently and the full 
single logarithmic result. 

In the figure above we plot the behaviour of the Monte Carlo estimate 
$S(t)$, the all-order, large $N_c$ resummed non-global contribution, versus the single-logarithmic evolution variable $t = \int_\epsilon^Q \frac{dk_t}{k_t}\alpha_s(k_t)$, for different geometries into which the emission is restricted (by direct measurement) such as a rapidity slice or a patch in the $\eta,\phi$ plane where $\phi$ is an azimuthal angle. The Sterman-Weinberg cross-section for narrow cones is essentially the problem of an infinite rapidity slice in which the emission is restricted to be less energetic than than $\epsilon$.
 \subsection{Buffer dynamics}

The most noteworthy feature of the result obtained for $S(t)$, the all orders resummed non-global (large $N_c$) contribution is the fact that at intermediate to large values of $t$ the behaviour is essentially a geometry independent (apart from overall normalisation) 
suppression which for $t \leq 0.5$ is a Gaussian fall of $e^{-k t^2}$.
To explain this geometry independence it has been suggested \cite{DS2} 
that the suppression at scale $t$ in a given region $\Omega$ results from a suppression at some earlier scale 
$t'$ in a buffer region surrounding $\Omega$ whose size gets bigger with decreasing $t'$. 
Due to this suppression mechanism far from $\Omega$ one becomes insensitive to the details of $\Omega$ itself ie whether it is a square patch in $\eta,\phi$ or a slice in $\eta$ etc.
\section{Phenomenological impact}
\begin{figure}[htbp]
\epsfxsize=8cm
\centerline{\epsfbox{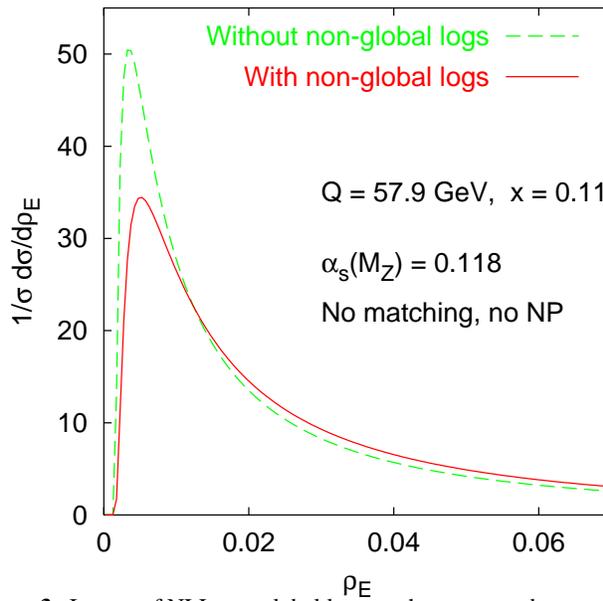}}
\caption{Impact of NLL non global logs on the resummed current-jet--mass distribution.}
\label{fig:good}
\end{figure}

Novel dynamics aside non-global logs are expected to (and do) make a sizable contribution to many observables and are therefore crucial in  a 
phenomenological sense. While the Sterman Weinberg jet fraction is a somewhat defunct quantity for various other reasons, resummations based on cone-type jet algorithms are candidates for non global logarithms. For example the invariant mass $(p_1+p_2)^2$ of dijets, near threshold, in say hadron-hadron production is a non global observable at single-log level if one defines the jets by cones around the leading hard partons (see e.g \cite{SKO}).

Most but not all 
current hemisphere event shape variables in DIS are non-global in nature due to the restriction to a single hemisphere. Here the non global single logs are classified as next-to--leading (NLL) since soft-collinear singularities (double logs) are also present. Inspite of being NLL the impact of non-global logs is evident from Fig.~\ref{fig:good}. One can see that for the current jet-mass the height of the peak is increased by approximately 50\% by not considering non-global logs.

Additionally there are variables like the $E_T$ flow away from jets 
where the non global logs are leading logarithmic (of the same order as the leading single logs generated by independent emission ). Their absence in that case was shown to have a large impact both qualitatively and quantitatively on the $E_T$ flow distribution \cite{DS2}. 
\section{Damage assessment, limitation and control}
The discovery of non global logs and the difficulties apparent in  achieving complete analytical insight (as opposed to large $N_c$ numerical estimates) regarding them 
led to the introduction of damage control measures to try and minimise the impact of the non global piece. Appleby and Seymour discovered that defining rapidity gaps in terms of minijet energies rather than hadronic energy affected the non global log piece. This was because the clustering of soft hadrons into minijets 
had the impact of pulling softest emissions inside rapidity gaps, outside the gap region, through clustering with  harder emitter. While this procedure did not eliminate non-global logs as there were configurations that survived the clustering, the effect  was somewhat reduced (see \cite{Aseymour}). 

Berger, Ku'cs, and Sterman, on the other hand, undertook the task of cleaning up the region outside the gap of relatively energetic emissions (compared to the gap energy)\cite{BKS}. This was done by using a jet-shape variable to squeeze jets and force them to be sufficiently narrow. This means that non global logs now appear as $\ln \frac{V}{\epsilon}$ where $V$ is the event shape value and $\epsilon$ the gap energy. 
If one chooses $V \sim \epsilon$ one finds that non global logs are suppressed. 
Dokshitzer and Marchesini \cite{DM} proceeded further with the analysis of such event-shape/energy-flow correlations. They were able to treat the general case where $V \gg \epsilon$ and showed a factorisation between resummed 
global $\alpha_s \ln^2V, \alpha_s \ln V$ terms and non global pieces involving powers of $\alpha_s \ln V/\epsilon$

\section{Conclusions}
The discovery of the non-global nature of several observables is certainly a significant one in terms of both theoretical advances and phenomenology. As regards the former it was generally believed that the NLL accuracy achieved using independent emission (coherent branching/angular ordering) was sufficient for a wide number of QCD observables and methods employing angular ordering were being widely used to claim single-log accuracy where in fact, in a number of cases, non 
global considerations were needed. Apart from several instances of mistaken omissions in the literature, non global logs themselves are of great intrinsic interest. The novel buffer dynamics that emerges from a study of the non-global piece is a good example of the complex nature of full QCD coherence and further investigation is required to better understand some of the gross features that have so far been presented \cite{DS2,BMS}.

On the phenomenology side given the sizable quantitative impact of non global logs they would be crucial to include in many 
studies ranging from fitting the strong coupling using event shapes to understanding the underlying event contribution in hadron-hadron collisions. 
The study of non global logs has also resulted in the introduction of new observables such as the event-shape/energy flow correlation which would be additionally interesting from the point of view of testing ideas on 
non-perturbative power corrections amongst many other studies.


\begin{thebibliography}{99}

\bibitem{DS1} 
M.~Dasgupta and G.P.~Salam, Phys.~Lett.~B 512, 323, (01).

\bibitem{DS2}
M.~Dasgupta and G.P.~Salam, JHEP 0208, 032, (02).

\bibitem{SW}
G.~Sterman and S.~Weinberg, Phys.~Rev.~Lett., 39, 1436, (77).

\bibitem{BMS} 
A.~Banfi, G.~Marchesini and G.E.~Smye, JHEP 0208, 006, (02).
\bibitem{SKO}
N.~Kidonakis, G.~Oderda and G.~Sterman, Nucl.~Phys.~B 531, 365, (02).
\bibitem{Aseymour}
R.B.~Appleby and M.H.~Seymour, JHEP 0212, 063, (02).

\bibitem{BKS}
C.F.~Berger, T.~Kucs and G.~Sterman, hep-ph/0303051.

\bibitem{DM}
Yu.L.~Dokshitzer and G.~Marchesini, hep-ph/0303101.

\end{thebibliography}
\end{document}